\documentclass[iop,apj]{emulateapj} \usepackage{apjfonts} \newif\ifdraft \drafttrue \newif\ifpre \pretrue
\usepackage{graphicx}
\usepackage{color}

\bibpunct{(}{)}{,}{a}{}{,}
\makeindex
\citeindextrue

\newcommand{\getlength}[1]{\ifx#1\end \let\next=\relax
            \else\advance\count255 by1 \let\next=\getlength\fi \next}
\newcount\switch
\newcommand{\Endmat}{\ifnum\switch=0$\fi}
\newcommand{\ifnularg}[1]{ \count255=0 \getlength#1\end \ifnum\count255=0 }
\newcommand{\ifm}{\makebox{}\ifmmode}

\long\def\ifundefined#1#2#3{\expandafter\ifx\csname
  #1\endcsname\relax#2\else#3\fi}
\newcommand{\beq}   { \begin{eqnarray} }
\newcommand{\eeq}[1]{ \ifnularg{#1} end{eanarray} \else
                      \label{#1}\end{eqnarray}    \fi }
\newcommand{\eeqn}{\nonumber\end{eqnarray}}
\newcommand{\ntab}[2]{ \multicolumn{1}{#1}{#2} }
\newcommand{\nntab}[2]{ \multicolumn{2}{#1}{#2} }
\newcommand{\nnntab}[2]{ \multicolumn{3}{#1}{#2} }

\newcommand{\ha}{\hphantom{a}}
\newcommand{\haa}{\hphantom{aa}}

\newcommand{\vex}{\vspace{1ex}}

\newcommand{\PIMA}{$\cal P\hspace{-0.067em}I\hspace{-0.067em}M\hspace{-0.067em}A$ }
\newcommand{\qcal}{\mbox{QCAL-1}}

\newcommand{\Number}[1]{\ifnum#1<10\relax0\number#1\else\number#1\fi}
\newcommand{\isodate}{
\count151=\time
\divide\count151 by 60
\count151=\count151
\multiply\count151 by 60
\count152=\time
\advance\count152 by -\count151
\divide\count151 by 60
\count152=\count151
\multiply\count151 by 60
\count153=\time
\advance\count153 by -\count151
\Number{\year}.\Number{\month}.\Number{\day}--\Number{\count152}:\Number{\count153}
}
\definecolor{Dred}{rgb}{0.312,0.070,0.070}
\definecolor{Dblue}{rgb}{0.070,0.070,0.312}
\definecolor{Dgreen}{rgb}{0.070,0.312,0.070}
\definecolor{Db}{rgb}    {0.050,0.0,0.320}

\newcommand{\Blb}[1]{\textcolor{Dblue}{\bf #1}}

\ifpre
\fi
\ifdraft
    \newcommand{\web}[1]{\Blb{\url{#1}}}
  \else
    \newcommand{\web}[1]{\url{#1}}
\fi
\newcounter{note}
\let\oldmarginpar\marginpar
\renewcommand\marginpar[1]{\-\oldmarginpar[\raggedleft\footnotesize #1]%
{\raggedright\footnotesize #1}}
\newcommand{\Note}[1]{\Rdb{#1}{\addtocounter{note}{1}%
\marginpar{\small\underline{\Rdb{Corr \arabic{note}}}}}}

\newcommand{\pz}{\phantom{0}}

\renewcommand{\Note}[1]{#1}

\shorttitle{The QCAL-1 KVN 43~GHz calibrator survey}
\shortauthors{Petrov et al.}
\ifpre
  \submitted{}
  \received{2012, July 24}
  \revised{2012, August 30}
  \accepted{2012, September 04}
\else
\fi

\begin{document}
\title{EARLY SCIENCE WITH KOREAN VLBI NETWORK: THE QCAL-1 43~GHz 
       CALIBRATOR SURVEY}

\author{
Leonid Petrov\altaffilmark{1},      
Sang-Sung Lee\altaffilmark{2},      
Jongsoo Kim\altaffilmark{2},        
Taehyun Jung\altaffilmark{2},       
Junghwan Oh\altaffilmark{2,3},      
Bong Won Sohn\altaffilmark{2,3},    
Do-Young Byun\altaffilmark{2,3},    
Moon-Hee Chung\altaffilmark{2},     
Do-Heung Je\altaffilmark{2},        
Seog-Oh Wi\altaffilmark{2},         
Min-Gyu Song\altaffilmark{2},       
Jiman Kang\altaffilmark{2},         
Seog-Tae Han\altaffilmark{2,3},     
Jung-Won Lee\altaffilmark{2},       
Bong Gyu Kim\altaffilmark{2,3},     
Hyunsoo Chung\altaffilmark{2}, and  
Hyun-Goo Kim\altaffilmark{2},       
}
\email{{\rm Correspondence send to} sslee@kasi.re.kr}
\altaffiltext{1}{Astrogeo Center, Falls Church, VA 22043, USA}

\altaffiltext{2}{Korean VLBI Network,
Korea Astronomy and Space Science Institute, 
776, Daedeokdae-ro, Yuseong-gu, Daejeon 305-348, 
Republic of Korea}

\altaffiltext{3}{Yonsei University Observatory, Yonsei University,
Seongsan-ro 50, Yonsei-ro, Seodaemun-gu, Seoul 120-749, Republic of Korea}

\ifdraft
\fi

\begin{abstract}


  This paper presents the catalog of correlated flux densities in
three ranges of baseline projection lengths of 637~sources from a 43~GHz 
($Q$-band) survey observed with the Korean VLBI Network. Of them, 14 objects
used as calibrators were previously observed, but 623 sources 
have not been observed before at $Q-$band with very long baseline interferometry
(VLBI). The goal of this work 
in the early science phase of the new VLBI array is twofold:
to evaluate the performance of the new instrument that operates 
in a frequency range of 22--129~GHz and to build a list of objects that 
can be used as targets and as calibrators. We have observed the list of 
799~target sources with declinations down to $-40\degr$. Among them, 
724 were observed before with VLBI at 22~GHz and had correlated flux densities
greater than 200~mJy. The overall detection rate is 78\%. The detection
limit, defined as the minimum flux density for a source to be detected 
with 90\% probability in a single observation, was in a range of 115--180~mJy 
depending on declination. However, some sources as weak as 70~mJy have 
been detected. Of 623 detected sources, 33 objects are detected for the first 
time in VLBI mode. We determined their coordinates with the median formal 
uncertainty 20~mas. The results of this work set the basis for future efforts 
to build the complete flux-limited sample of extragalactic sources at frequencies 
22~GHz and higher at 3/4 of the celestial sphere.

\end{abstract}


\section{Introduction}
\label{s:introduction}

  The Korean VLBI Network (KVN) is the first dedicated very long naseline 
interferometry (VLBI) network in East Asia in millimeter wavelengths.
The KVN was built by Korea Astronomy and Space Science Institute (KASI)
in order to achieve the following major goals: (1)~to study the formation and 
death of stars with observing water (H${}_2$O), methanol (CH${}_3$OH),
and silicon monoxide (SiO) masers at high resolutions, (2)~to investigate the 
structure and dynamics of our own Galaxy by conducting highly accurate 
astrometric VLBI observations of the galactic radio sources, 
and (3)~to study the nature of active galactic nuclei (AGNs) and their 
population at high frequencies. The KVN as a dedicated VLBI network also 
aims to study the spectral and temporal properties of transient sources 
such as bursting star-forming regions, intra-day variable compact radio 
sources, gamma-ray flaring AGNs, and other objects by conducting systematic 
multi-wavelength monitoring campaigns \citep{r:kim+04,r:lee+11}

\begin{figure}[h]
  \ifpre \begin{center} \fi
  \caption{
           Korean VLBI network.
  }
  \label{f:kvn_map}
  \par\medskip\par
  \includegraphics[width=0.46\textwidth]{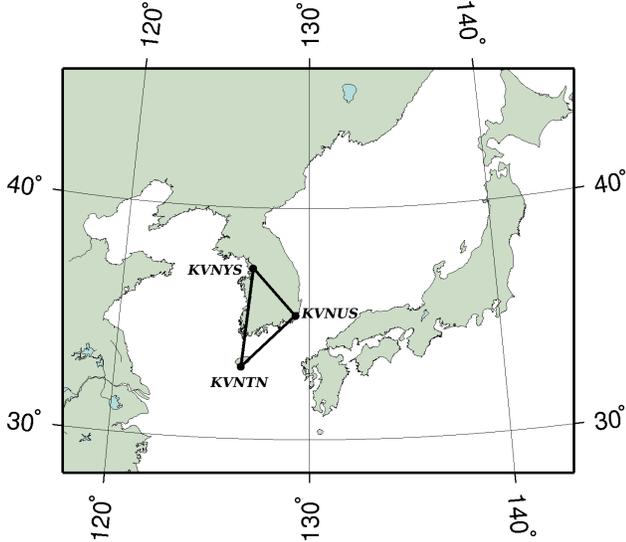}
  \ifpre \end{center} \fi
\end{figure}

The KVN consists of three 21\ m radio telescopes: in Seoul, KVN Yonsei Radio 
Telescope (KVNYS), in Ulsan, KVN Ulsan Radio Telescope (KVNUS); and 
in Jeju island, Korea, KVN Tamna Radio Telescope (KVNTN).
The baseline lengths are in a range of 305--476~km (see Figure\ref{f:kvn_map}).
All antennas have an identical design. The aggregate root mean square 
(rms) deviation of the antenna surface from a paraboloid is 0.12~mm, 
which allows us to observe at frequencies up to 150~GHz. The antennas are 
equipped with the quasi-optic system that allows simultaneous observations 
at 22, 43, 86, and 129~GHz. This system is described in detail
in \citet{han+08}.

By 2011, the 22 and 43~GHz receivers were installed and carefully tested.
The dual beams at two frequencies are well aligned within 
$5''$. The pointing errors are less than $3''$ in azimuth and 
elevation. The measured aperture efficiencies are greater than 64\% 
at 22~GHz and greater than 62\% at 43~GHz. More detailed results of 
performance tests of the antenna and receivers in the single-dish mode 
are presented in \citet{r:lee+11}.  Receivers at 86 and 129~GHz will be 
tested in 2012.

The signals digitized by the samplers in the receiver room are processed
by the KVN Data Acquisition System (DAS) to get spectra for single-dish
spectroscopy observations. In VLBI operation, the digitized signals processed
in the DAS are recorded onto disks using Mark-5B. In 2011 the aggregate
recording rate was limited by 1024~Mbps that can be used either entirely
for one band or split between bands.

The KASI has all hardware and software to support {\it the full cycle 
of VLBI observations} either as a three-element interferometer or as a part 
of a larger VLBI network: scheduling, antenna control system, data recording, 
correlation, post-correlation data processing, astrometry, geodesy, and imaging 
analysis.First fringes between KVN stations at 22~GHz in a recording rate
of 1~Gbps were obtained on 2010 June 8. Test observations on 
2010 September 30 confirmed that the VLBI system works according 
to specifications. Detailed analysis of system performance will be given 
in S.~Lee et al. (2013 in preparation).

  These first tests allowed us to start an early science program including 
multi-frequency phase referencing observations. The feasibility of the 
multi-frequency phase referencing observations strongly depends on the 
information about phase calibrators such as their structures, flux densities, 
global distribution, etc \citep{r:jung11}. Therefore it is necessary 
to establish a catalog of radio compact sources at high frequency bands with 
large VLBI surveys. This determined our choice for the first early science 
project for VLBI observations in continuum mode.

In this paper we present results of the 43~GHz VLBI survey conducted
at the three-element KVN radio interferometer. We describe the strategy for
source selection, observations, data analysis and present the catalog of
correlated flux densities. Finally, we summarize main findings and
make conclusions about performance of the KVN for AGNs studies.

\section{The \qcal\ survey: goals and source selection}
\label{s:goals}

  Although the accumulative radio fundamental 
catalog\footnote{Available at \web{http://astrogeo.org/rfc}} (L.~Petrov \&
Y.~Kovalev 2013, in preparation) that includes all the sources observed with VLBI 
in a survey mode by 2012 August has 7216 objects, the majority of them were 
observed at frequencies 8~GHz and lower. This catalog is based on Very Long
Baseline Array (VLBA) Calibrator Survey 
\citep{r:vcs1,r:vcs2,r:vcs3,r:vcs4,r:vcs5,r:vcs6}
and other large surveys. There were only several VLBI surveys at frequencies 
higher than 24~GHz. \citet{r:kq} observed 132 sources at $Q$-band in 2002--2003, 
\citet{r:bu2} run an ongoing project of monitoring 34~AGNs since 2007,
and \citet{r:ka} started to work on the catalog of $\sim\!300$ objects
at 32~GHz that will become available in the future. Considering that 
13~sources overlap in both projects, the total number of sources with 
known  $Q$-band flux densities at milliarcsecond scales with their brightness 
distributions publicly available is 153. This count does not include 
individual sources that were observed as targets under other programs. 
All these sources have the correlated flux density at baselines longer 
than 900~km, which corresponds to 130~mega wavelengths, greater than 200~mJy.

  We would like to increase this list for two reasons. First, we need to have
a set of sources suitable as calibrators for high frequencies with KVN 
and VERA (VLBI Exploration of Radio Astrometry). Observations with quick 
switching between a weak target and a bright calibrator allow to extend 
the coherence time from several minutes to hours, provided a calibrator 
located within several degrees from a target can be detected. Second, 
construction of a flux-limited sample of sources at high frequencies provides 
an opportunity to study the population of AGNs. At frequency 43~GHz and higher 
the radio emission at parsec scales is dominated by the core. 
VLBI observations of a large flux-limited sample allow to derive a number 
of interesting statistics. In particular, using such a sample we can

\begin{itemize}
  \item investigate compactness determined as a ratio of the correlated 
        flux density at short and long baselines to the single-dish 
        flux density and compare it with compactness at 2.2, 8.4~GHz, 
        and 22~GHz.

  \item measure the brightness temperature for the core and jet components 
        in order to populate a model of the distribution of the observed core 
        brightness temperature in terms of the intrinsic properties of 
        relativistic jets such as the brightness temperature, bulk motion, 
        viewing angle. The data will also provide the unbiased sample needed 
        to investigate the unified scheme between quasars, BL Lacs objects, 
        and galaxies.

  \item analyze the relationship between the core optical depth 
        as measured by the spectral index and other properties of the 
        jet emission and optical properties.

  \item investigate the variation of the spectral index along 
        the presumably optically thin jets and comparison with the 
        conditions in small-scale and larger-scale jets.
\end{itemize}

  To commence building a large sample of compact sources at 43~GHz, we
analyzed prior VLBI surveys at 22--24~GHz: the $K$/$Q$ VLBA survey 
\citep{r:kq}, the VERA 22~GHz fringe search survey \citep{r:vera_fss},
the VLBA Galactic Plane Survey \citep{r:vgaps}, the EVN Galactic Plane 
Survey \citep{r:egaps}, and the VERA $K$-band Calibrator Survey \citep{r:kcal}.
We selected 878 sources in these surveys that have correlated flux densities
at scales less than 10~mas exceeding 200~mJy. After subtracting 153 sources
that have already been observed in $Q$-band surveys, we got the list of 
725 target sources. We added to this list two samples of the sources that
were not previously observed with VLBI which we could reasonably expect
to be bright enough for being detected at $Q$-band with the KVN. These
additional sources included 49 objects from the CGRaBS catalog
\citep{r:cgrabs}, many of them are part of the OVRO 40~m telescope 
monitoring program at 15~GHz \citep{r:ovro} (other 1576 objects have 
already been observed with VLBI) and 26 sources from the AT20G 
catalog \citep{r:at20g} with spectral index ($S \sim f^{\alpha}$)
greater than $+0.5$ that have the flux density extrapolated to 43~GHz 
greater than 200~mJy. In total, the list of targets included 799~objects.

\subsection{Observations}
\label{s:observations}

  We observed target sources in three observing sessions in December 18, 19,
and 22, 2011. The first two sessions had durations 28 and 27 hr
respectively. In the 50 minutes long third session we observed the sources 
missed in the first two sessions. Each source was scheduled in one 
scan, 90~s long, in the first observing session and in one scan in the
second observing session. The algorithm for automatic scheduling implemented 
in software sur\_sked\footnote{\web{http://astrogeo.org/sur\_sked}}
selected the sources in a sequence that minimizes slewing. 

  In addition to target sources, every half an hour a source from the 
list of 34~blazars that are regularly monitored by the Boston group 
\citep{r:bu1,r:bu2} was inserted. The scheduling algorithm has picked up 14 
blazars. These bright sources were used for bandpass calibration and 
for evaluation of station-dependent gain corrections.

  Every hour a tipping curve was measured: each antenna recorded system
temperature at eight different elevations. These measurements allow us 
to monitor changes in the opacity of the atmosphere.
 
  The left circular polarization data were recorded using Mark-5B units 
with 2~bit sampling within 496~MHz wide band in a range of 42.850--43.346~GHz 
in 16~intermediate sub-bands of 16~MHz wide, equally spaced with a step 
of 16~MHz using a digital filter. The aggregate recording rate was 1024~Mbps.

\section{Data analysis}
\label{s:anal}

  The data were correlated with DiFX-2 correlator \citep{r:difx2} with 
spectral resolution of 0.125~MHz and accumulation periods of 1~s long. 
The scheme of post-correlator data analysis was similar to the analysis 
of KCAL VERA calibrator survey at 22~GHz \citep{r:kcal}. First, 
for each observation we found such phase delay rates, group delays, and 
group delay rates that after coherent averaging of complex cross-correlation 
samples rotated according to these parameters the fringe amplitude of the sum
reached the maxima. We used analysis software 
\PIMA\footnote{ \web{http://astrogeo.org/pima} } for performing this task. 
A dataset that originates from processing data of one 
scan at one baseline we call an observation.

  Analyzing the distribution of achieved signal-to-noise ratios (S/Ns), 
we found the probability of false detection for observations with 
no signal, but only noise is less than 0.01 when the S/N is greater than 5.4. 
We used this value of the S/N as a threshold for a preliminary screening 
of detected and not detected observations.

  Then total group delays and delay rates as well as some auxiliary information
was written into a database for further processing with VTD/post-Solve 
software\footnote{ \web{http://astrogeo.org/vtd} }
for analysis of absolute astrometry and geodesy observations. We solved 
for coordinates of new sources, baseline lengths, clock function and residual 
atmosphere path delay in zenith direction modeled with a linear B-spline with 
spans of 1~hr for every station. This was done for two purposes. First, we 
were able to adjust positions of new sources and improve a~priori baseline 
lengths. Second, considering that for a majority of sources their positions 
are known at a milliarcsecond level, we can make an additional test whether 
a given source was detected. The group delay search window was 8 $\mu$s. 
If a source is not detected, its estimate of group delay will be uniformly 
distributed in a range of $\pm 4 \mu$s. The distribution of post-fit 
residuals of detected sources has the weighted root mean squares (wrms) 
55--80~ps. That is why a non-detected source will show up as 
an outlier. The probability that a non-detected source which position is 
not adjusted will accidentally have the post-fit residual \Note{greater 
than 5 times wrms, i.e., $\pm 0.4$~ns}, and therefore, will not be marked 
as an outlier, is $10^{-4}$.

  \Note{After additional screening for non-detections by analyzing 
residuals of parameter estimation and elimination of observations with 
residuals exceeding five times their wrms we performed the amplitude 
calibration.} Previous measurements of antenna gains 
showed that they vary with elevation within 5\%--8\%. Table~\ref{t:gain} 
provides parameters of gains used in our analysis.

\begin{table}[ht!]
   \caption{ \ifpre \rm \fi
           Dependence of Gain at 43~GHz for KVN Antennas Measured 
           in 2011 February--March. 
           }                       
   \label{t:gain}
   \begin{tabular}{rrrrr}
      Station & DPFU    & \ntab{c}{$A_0$} & \ntab{c}{$A_1$} & \ntab{c}{$A_2$}           \\
      \hline
      KVNTN   &  0.0747 & 0.979990 &  $1.23772\cdot10^{-3}$ &  $-1.91393 \cdot 10^{-5}$ \\
      KVNUS   &  0.0714 & 0.974197 &  $1.41639\cdot10^{-3}$ &  $-1.94371 \cdot 10^{-5}$ \\
      KVNYS   &  0.0850 & 0.923354 &  $3.47852\cdot10^{-3}$ &  $-3.94673 \cdot 10^{-5}$ \\
      \hline
   \end{tabular}
   \tablecomments{Gain in Jy/K is expressed as 
           DPFU$ \cdot (A_0 \: + \: A_1 \cdot e \: + \: A_2 \cdot e^2)$, where
           $e$ is elevation in degrees.}
\end{table}

  Analysis of system temperature measurements revealed small variations with
time and with elevation angle. The measured system temperature is
considered as a sum of three terms: the receiver temperature $T_{\rm rec}$,
the spillover temperature $T_{\rm spill}$, and the contribution of 
the atmosphere:
\begin{eqnarray}
   T_{\rm sys} = T_{\rm rec} \: + \: 
                 \eta \, T_{\rm atm} [ 1 - e^{-\beta \, m(e)} ] 
                 \: + \: (1 - \eta)T_{\rm spill}
                 ,
   \label{e:e3}
\end{eqnarray}
   where $T_{\rm atm}$ is the average temperature of the atmosphere;
$\beta$ is the atmosphere opacity; $\eta$ is the spillover efficiency,
which includes rear spillover, scattering, blockage, and ohmic loss 
efficiency; and $m(e)$ is the wet mapping function: the ratio of the 
non-hydrostatic constituent of the path delay in the neutral atmosphere 
at the elevation $e$ to that path delay in the zenith direction. Since 
the observations were made at different elevation angles down 
to $10\degr$, decomposition of system temperature according 
to expression~\ref{e:e3} can be easily done using non-linear least squares. 
Sky tipping curve analysis procedure of the KVN follows the way described 
in \citet{r:kp-manual}, Section~6.5.1. \Note{We used NMFw wet 
mapping function \citet{r:nmf} in our work}. Estimates of the opacity, spillover 
efficiency, and receiver temperature from system temperature measured 
during observations were within 5\%--10\% from results derived from tipping 
curves. Variations of both receiver temperature and opacity varied within 
10\% over four days. Table~\ref{t:sefd} shows average receiver temperature, 
opacity, and system equivalent flux density (SEFD) in zenith direction during 
observations. Using estimates of opacity, we reduced our system temperature 
to the top of the atmosphere and calibrated fringed amplitudes by multiplying 
them by a factor $\sqrt{T_{\rm sys,1}\, T_{\rm sys,2}}/G$, where $G$ is 
the a~priori gain.

  Since each target source was scheduled in two scans of a three-element
interferometer, the self-calibration method for imaging will not work 
because too few data are available. We computed predicted correlated flux 
densities for every observation of 14~amplitude calibrator sources using 
their brightness distributions in the form of CLEAN components from VLBA 
observations under the Boston blazar monitoring program. We processed 
images for epoch 2011 December~02 or 2012 January~27, i.e., within one month 
from our observations. Using these
predicted correlated flux densities, and comparing them with the calibrated
fringe amplitudes from our observations, we estimated average multiplicative 
gain factors using least squares. These factors were considered constant 
over each individual observing session. Detailed discussion of this method and 
approaches for evaluation of errors associated with this method are given 
in \citet{r:kcal}. Using the same technique, we found the calibration errors
of the \qcal\  survey were within 15\%.

\begin{table}[h]
   \caption{\ifpre \rm \fi
            Average Receiver Temperature, Opacity, the System
            Equivalent Flux Densities (SEFDs) in Zenith Direction,
            and a~Posteriori Multiplicative Gain Corrections 
            During the Campaign.}
   \label{t:sefd}
   \ifpre \par\vspace{-2ex} \begin{center} \fi  
   \begin{tabular}{l r r r r r}
      Station & $T_{\rm rec}$ (K) & $\eta$ \ha & $\beta$ \ha & SEFD (Jy) & Gain${}_{\rm corr}$ \\
      \hline
      KVNTN   &     89 \haa & 0.945 & 0.081 & 2080 \haa & 1.10 \haa \\
      KVNUS   &     51 \haa & 0.925 & 0.082 & 1460 \haa & 1.06 \haa \\
      KVNYS   &    103 \haa & 0.945 & 0.103 & 2810 \haa & 1.31 \haa \\
      \hline
   \end{tabular}
   \ifpre \end{center} \fi
   \tablecomments{ 
                  Although receiver design for all three stations 
                  is identical, the thermal isolator of Ulsan  $Q$-band receiver
                  is different and it contributes to a significant reduction 
                  of receiver temperature. Upgrade of thermal isolators 
                  at KVNYS and KVNTN is planned in near future.
                 }
\end{table}

\section{\qcal\  Catalog}
\label{s:catalog}

\begin{table*}[ht!]
  \caption{\ifpre \rm \fi
           The First 8 Rows of the Catalog of Correlated Flux Densities 
           of 637 Sources that have at Least Three Detections in 
           KVN \qcal\  Observing Campaign.
          }
  \label{t:flux}
  \ifpre \vspace{-7ex} \begin{center}\fi
  \begin{tabular}{ c l c c r r @{\qquad} r r r r r r @{\qquad} l l }
         \multicolumn{12}{c}{ } \vspace{3ex} \\
         \multicolumn{2}{c}{\normalsize Source names}  &
         &
         \multicolumn{1}{c}{\normalsize Stat.}    &
         \multicolumn{3}{c}{\normalsize Corr. flux density}    &
         \multicolumn{3}{c}{\normalsize Errors of $F_{\rm corr}$}  &
         \multicolumn{2}{c}{\normalsize Source coordinates}     
         \\ [0.5ex]
         \ntab{c}{(1)}    &
         \ntab{c}{(2)}    &
         \ntab{c}{(3)}    &
         \ntab{c}{(4)}    &
         \ntab{c}{(5)}    &
         \ntab{c}{(6) }   &
         \ntab{c}{(7)}    &
         \ntab{c}{(8)}    &
         \ntab{c}{(9)}    &
         \ntab{c}{(10)}   &
         \ntab{c}{(11)}   &
         \ntab{c}{(12)}
         \vspace{1ex}
         \\ 
         IAU name     &
         IVS name     &
         flag         &
         \#Det        &
         $F_{<21}$    &
         $F_{21-42}$  &
         $F_{>42}$    &
         $E_{<21}$    &
         $E_{21-42}$  &
         $E_{>42}$    &
         Right ascen  &
         Declination
         \\ 
                       &
                       &
                       &
                       &
         \ntab{c}{Jy}  &
         \ntab{c}{Jy}  &
         \ntab{c}{Jy}  &
         \ntab{c}{Jy}  &
         \ntab{c}{Jy}  &
         \ntab{c}{Jy}  &
         \multicolumn{1}{l}{\pz h \pz m \pz s} &
         \multicolumn{1}{l}{\pz\pz $\degr \pz\pz {}' \pz\pz {}''$} \\
      \hline
      J0013$+$4051 & 0010$+$405 &   & 6 & -1.000 & -1.000 & 0.612 & -1.000 & -1.000 & 0.018 & 00 13 31.1302 & $+$40 51 37.144 \\
      J0013$-$0423 & 0011$-$046 &   & 3 & -1.000 & -1.000 & 0.087 & -1.000 & -1.000 & 0.014 & 00 13 54.1309 & $-$04 23 52.294 \\
      J0014$+$6117 & 0012$+$610 &   & 6 & -1.000 & -1.000 & 0.281 & -1.000 & -1.000 & 0.016 & 00 14 48.7921 & $+$61 17 43.542 \\
      J0017$+$8135 & 0014$+$813 &   & 7 & -1.000 &  0.257 & 0.254 & -1.000 &  0.016 & 0.020 & 00 17 08.4749 & $+$81 35 08.136 \\
      J0022$+$0608 & 0019$+$058 &   & 6 & -1.000 &  0.449 & 0.448 & -1.000 &  0.016 & 0.017 & 00 22 32.4412 & $+$06 08 04.268 \\
      J0027$+$5958 & 0024$+$597 &   & 6 & -1.000 & -1.000 & 0.189 & -1.000 & -1.000 & 0.016 & 00 27 03.2862 & $+$59 58 52.959 \\
      J0029$+$0554 & 0027$+$056 &   & 6 & -1.000 &  0.295 & 0.261 & -1.000 &  0.015 & 0.015 & 00 29 45.8963 & $+$05 54 40.712 \\
      J0038$+$1856 & 0035$+$186 & N & 6 & -1.000 &  0.099 & 0.086 & -1.000 &  0.014 & 0.015 & 00 38 28.8915 & $+$18 56 17.641 \\
      \hline
  \end{tabular}
  \tablecomments{Units of right ascension are hours, minutes and seconds. 
                 Units of declination are degrees, minutes and seconds.
                \hspace{10em}\hfill\linebreak
                (This table is available in its entirety in machine-readable
                 and Virtual Observatory (VO) forms in the online journal.
                 A portion is shown here for guidance regarding its form
                 and content.)
                }
  \ifpre \end{center} \fi
\end{table*}

  Among 813 observed sources, 799 targets and 14 calibrators, 637 sources
were detected in three or more observations, 27 objects were detected in 
two observations, 20 objects were detected in only one observation, and
129 objects were not detected at all. If a source had a residual by modulo
greater
than 0.4~ns in the astrometry solution, \Note{we counted that observation as 
a non-detection, regadrless of its S/N}. We consider a source reliably detected if it was detected in
three or more \Note{observations}. Since the probability of false detection 
of an individual observation is 0.01, and considering such events are 
independent, the probability that all three detections with S/N $> 5.4$ 
are spurious is $10^{-6}$. The probability that falsely detected observations 
have residuals less than 0.4~ns for all three observations is $10^{-12}$ 
if to regard these cases as statistically independent and $10^{-4}$ 
if to regard them as totally dependent. 

  For each 637 detected sources we computed the median correlated flux density
in three ranges of baseline projection lengths: [0, 21], [21, 42],
and [42, 68] megawavelengths, which corresponds to the ranges of 
150, 300, and 477~km. The first 8 rows of the catalog are presented 
in Table~\ref{t:flux}.

  Errors in estimates of \Note{correlated flux density} are determined by two factors:
errors in calibration and unaccounted source structure. Since the majority
of sources are barely resolved at 43~GHz with angular resolution 5~mas that
we had in our observations, the first factor dominates. As a test, we analyzed
the first and the second observing session separately. Comparison of flux 
density estimates from independent experiments showed us that they are 
consistent within 5\%--10\%. \Note{This indicates that our estimate of 
the calibration uncertainty 15\% is close rather to an upper limit of 
actual errors than the low limit}. However, due to scarceness of redundant 
observations we have to refrain from attempts to derive more robust 
statistics of flux density measurements.



  In our astrometric solution that used all three observing sessions 
the following parameters were estimated with least squares: positions of 33 new 
sources, baseline lengths, and the nuisance parameters, such as clock 
function and residual atmosphere path delay in zenith direction. We kept 
positions of remaining 604 sources fixed to their values from the recent 
update of the Radio Fundamental Catalog that used all available VLBI
observations under geodesy and absolute astrometry programs since 1980 April 
through 2012 June. Positions of these 604 sources defined the orientation 
of the new catalog of 33 sources. 

\begin{table*}[ht!]
   \caption{\ifpre \rm \fi
            The first 8 rows of the \qcal\  catalog of positions
            of 33 sources never before observed with VLBI.
           }
   \label{t:cat}
   \ifpre \par\vspace{-12ex} \begin{center} \fi
   \begin{tabular}{ c l l l r r r r r }
         & & & & & & & &
         \vex \vex \vex \\
         & & & & & & & & 
         \vex \\
         \nntab{c}{Source Names}      &
         \nntab{c}{J2000 Coordinates} &
         \nnntab{c}{Errors (mas)}     &
         \nntab{c}{Stat}
         \vex \\
         \cline{1-2}
         \cline{3-4}
         \cline{5-7}
         \cline{8-9}
         \vex \\
         \ntab{c}{IAU  }   &
         \ntab{c}{IVS  }   &
         \ntab{c}{Right $\,$ ascension } &
         \ntab{c}{Declination   }   &
         \ntab{c}{$\Delta \alpha$ } &
         \ntab{c}{$\Delta \delta$ } &
         \ntab{c}{Corr  }  &
         \ntab{c}{\# Obs } &
         \ntab{c}{\# Exp } 
         \vex \\
         \ntab{c}{(1)}    &
         \ntab{c}{(2)}    &
         \ntab{c}{(3)}    &
         \ntab{c}{(4)}    &
         \ntab{c}{(5)}    &
         \ntab{c}{(6)}    &
         \ntab{c}{(7)}    &
         \ntab{c}{(8)}    &
         \ntab{r}{(9)}    
   \vex \\
   \hline
   J0038$+$1856 & 0035$+$186 & 00 38 28.8915 & $+$18 56 17.641 & 20.0 & 13.0 & $-$0.479 & 6 & 2 \\
   J0202$+$3943 & 0159$+$394 & 02 02 01.6563 & $+$39 43 21.556 & 11.0 &  5.8 & $-$0.235 & 6 & 2 \\
   J0204$+$4005 & 0201$+$398 & 02 04 05.1941 & $+$40 05 03.514 & 20.0 & 12.0 & $-$0.376 & 6 & 2 \\
   J0242$+$2653 & 0239$+$266 & 02 42 20.8303 & $+$26 53 37.722 & 30.0 & 17.0 & $-$0.092 & 4 & 2 \\
   J0251$+$7226 & 0246$+$722 & 02 51 37.3545 & $+$72 26 55.847 & 50.0 & 14.0 & $-$0.287 & 6 & 2 \\
   J0251$+$3734 & 0248$+$373 & 02 51 59.1686 & $+$37 34 18.177 & 33.0 & 23.0 & $-$0.097 & 4 & 2 \\
   J0452$+$1236 & 0449$+$125 & 04 52 42.6001 & $+$12 36 24.586 & 24.0 & 18.0 & $-$0.565 & 5 & 2 \\
   J0502$-$2057 & 0500$-$210 & 05 02 10.4053 & $-$20 57 16.590 & 30.0 & 31.0 & $-$0.615 & 3 & 1 \\
   \hline
   \end{tabular}
   \tablecomments{Units of right ascension are hours, minutes and seconds. 
                  Units of declination are degrees, minutes and seconds.
                 \hspace{10em}\hfill\linebreak
                 (This table is available in its entirety in machine-readable
                  and Virtual Observatory (VO) forms in the online journal.
                  A portion is shown here for guidance regarding its form
                  and content.)
                 }
   \ifpre \end{center} \fi
\end{table*}

  The first 8 entries of the catalog are presented in Table~\ref{t:cat}.
The semi-major axes of the error ellipse from formal uncertainties of 
source positions are in a range from 8 to 45~mas with the median value 
of 20~mas. In 2012 March and April, six sources from Table~\ref{t:cat} were
observed in one scan at 7.9--8.9~GHz with VLBA as fillers under ongoing
astrometry project for observations of a complete sample of Two Micron All
Sky Survey galaxies \citep{r:v2m}. Uncertainties of source position determined 
from these observations were in a range of 0.3--3~mas, more than one order 
of magnitude smaller than in \qcal\  observations. The average arc length 
between VLBA and KVN positions is 26~mas, which is consistent with their 
formal uncertainties.

\ifpre \vspace{2ex} \fi  

\section{Discussion}
\label{s:discussion}

  Comparing the correlated flux densities and the S/Ns, we can determine
for each observations with S/N $>$ 10 the minimal correlated flux density 
that corresponds to the S/N cutoff equal to 5.4. We built the cumulative 
distribution of the correlated flux densities that correspond to this cutoff
and analyzed them. Sources with low declinations, and we observed sources 
with declinations as low as $-41\degr$, are observed inevitably at low 
elevations through a thick layer of the atmosphere. This certainly affects 
our ability to detect a source. Table~\ref{t:det} shows the minimal correlated 
flux density for a source to be detected over 90~s integration time 
with a given probability derived from analysis of the cumulative distribution.

\begin{table}[ht!]
   \caption{\ifpre \rm \fi
            The Minimal Correlated Flux Density in Jy for a Source
            that has the Probabilities 50\%, 70\%, 90\%, and 95\% 
            of Detection in One Observation During \qcal\  Campaign.
           }
   \label{t:det}
   \ifpre \begin{center} \fi
   \begin{tabular}{r r r r r}
      Decl. Range               & 50\% & 70\% & 90\% & 95\% \\
      \hline
      $[-20\degr$, $+90\degr]$  &   75 &   90 &  115 & 130  \\
      $[-30\degr$, $-20\degr]$  &  105 &  115 &  130 & 140  \\
      $[-40\degr$, $-30\degr]$  &  140 &  160 &  180 & 190  \\
      \hline
   \end{tabular}
   \ifpre \end{center} \fi
\end{table}

\begin{figure}[ht!]
  \caption{
           Distribution of the correlated flux densities at baseline
           projection lengths longer than 21.44 megawavelengths. The last
           bin of the histogram has all the sources with correlated
           flux density $>1$ Jy.
  }
  \label{f:hist}
  \par\medskip\par
  \includegraphics[width=0.46\textwidth]{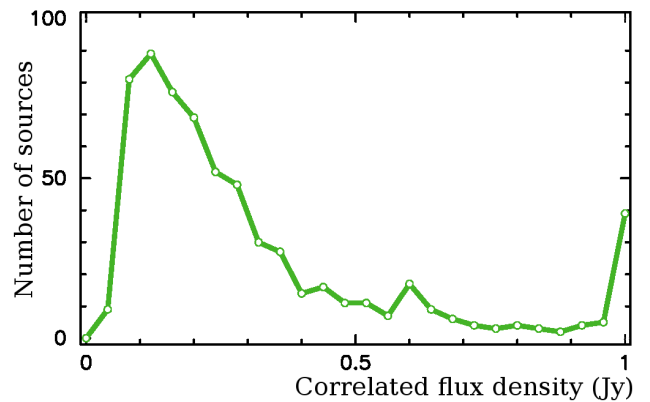}
\end{figure}

  The distribution of correlated flux densities from \qcal\  observations 
shows a growth of the number of detected sources with decreasing their
flux densities until it drops at around 110~mJy (Figure~\ref{f:hist}).
The flux density where the drop begins agrees well with our estimate 
of the detection limit presented in Table~\ref{t:det}. 

  The distribution does not include other sources from $K$/$Q$ and Boston
monitoring surveys. At present, we are not in a position to present evidence 
that the cumulative list of 776 objects from these two surveys and from 
\qcal\  is complete at a certain flux density level. More observations and 
meticulous completeness analysis of the parent source list are needed in 
order to draw a conclusion about completeness of the $Q$-band sample. 
At present, we can firmly say that at least 534 sources have correlated flux 
densities greater than 200~mJy at projection baseline length of 40~mega 
wavelengths and 381~objects have correlated flux densities greater than 
300~mJy at 3/4 of the celestial sphere with declination $>$ -30 $\deg$.

\Note{We can make a coarse estimate of the number of sources with the 
flux density at $Q$-band greater than 100~mJy. We computed the ratios
of  $Q$-band correlated flux density to $K$-band correlated flux density.
For 73\% sources the ratio exceeds 0.5 and for 22\% exceeds 1.0.
The dependence of the number of sources $N$ at $\delta > -30\degr$ 
on the $K$-band correlated flux density $S$ for $S>0.35$~Jy is 
approximated by $N = 182 \cdot S^{\:-1.279}$, where $S$ is 
in Jansky. Extrapolating this dependence to flux densities 0.2 and 
0.1~Jy, we get the expected number of sources: 1426 and 3460,
respectively. Considering that 73\% sources with $K$-band correlated 
flux density $>0.2$~Jy and 22\% sources with $K$-band correlated flux 
density in a range of $[0.1, 0.2]$~Jy can be detected at  $Q$-band 
with the KVN, we get a total expected number to detections: 
$\sim 1500$ objects.}

  We selected integration time 90~s for target sources. However, visual
inspection of variations of fringe amplitudes with time for strong sources
did not show a sign of decorrelation due to either the atmosphere or the 
frequency standards. We re-ran fringe fitting by dropping cross-correlation 
samples after 45 and after 60~s since the scan nominal start. If decorrelation 
due to phase fluctuations caused by the atmosphere path delay or by the 
frequency standards is negligible, the S/N ratio grows with 
an increase of averaging time as a square root of time. Deviation from this 
dependence in a form of $S/N(t_2) = S/N(t_1) \, D(t_2) \sqrt{t_2/t_1}$
provides us a measure of decorrelation $D(t)$. We found that the average 
decorrelation factor is 0.99 when the coherence time is increased 
from 45 to 60~s and 0.98 when the coherence time is increased from 60 to
90~s. This indicates that the coherence time during \qcal\  survey was 
significantly longer than 90~s.

  Considering that (1)~100\% of scheduled observations were observed,
correlated, and fringe fit; (2)~no noticeable decorrelation due to either 
frequency standard or atmosphere was found; (3)~system temperature was
measured for every scan with no abnormalities; (4)~receiver temperature
showed variations within 10\%; (5)~the median detection limit was 110~mJy;
(6)~positions of new sources were determined with median accuracies 20~mas,
we regard the early science KVN observations for \qcal\ project as fully
successful.

\section{Summary}
\label{s:summary}

  The major result is the catalog of flux densities of 637 compact 
sources at 43~GHz. Their errors do not exceed 15\%. The number of sources 
detected in  $Q$-band surveys grew by a factor of five and reached 776 objects. 
Integration time 90~s allowed us to detect a 115~mJy source at declination 
$> -20\degr$ with the probability 90\% in the winter season. We found that our 
observations were not limited by coherency time of either the atmosphere or 
the frequency standards. We may tentatively suggest that the coherence time 
could be increased to 180~s. We can conclude that the KVN is able to reliably 
detect at  $Q$-band sources with correlated flux densities greater than 
100~mJy, at least in winter time. More observations are needed in order 
to judge whether the coherence time during \qcal\  campaign was representative.

  We observed sources at elevations as low as $10\degr$ and with 
declinations as low as $-41\degr$. The detection limit for sources with 
declinations below $-30\degr$ was a factor of 1.5--2 worse that for sources
with declinations greater than $-20\degr$.

  We determined coordinates of 33~new sources that have not been observed before 
with VLBI. The median formal position uncertainty from two scans is 20~mas.
Comparison of these positions with positions of six sources derived from 
analysis of VLBA observations at 8~GHz in 2012 confirmed that the formal 
uncertainties are realistic. Observing sources in 12--15 scans which 
requires approximately 30~minutes per source will bring position uncertainty 
down to 5--10~mas. 

  The 56~hr long observing campaign is not sufficient to reach 
completeness of the AGN population at a certain flux density level. 
New campaign \mbox{QCAL-2} is planned to achieve this goal.
The on-going KVN $K$-band calibrator survey (J.~A.~Lee et al. 2013,
in preparation) is expected to increase the density of calibrators at 22~GHz
for phase referencing observations and it will be used as a pool of targets
for \mbox{QCAL-2}.


  These early science results met or exceeded our expectations of KVN 
performance for AGN studies. We conclude that the number of AGNs that the 
KVN is able to detect at  $Q$-band using 1024~Mbps recording rate is well over 
one thousand. These sources can also be used as phase calibrators for 
observing much weaker targets. With the use of \qcal\ results, the 
probability to find a calibrator for KVN observations within $2\degr$ 
of any target is 30\%. Future more deep surveys promise to increase 
significantly this probability and therefore, boost our ability to detect 
interesting targets at high frequencies.

\ifpre \par\bigskip\bigskip\par

\section{Acknowledgments}
\label{s:acknowledgments}

  We are grateful to all staffs and researchers of KVN who helped to develop and 
evaluate the KVN systems. This work was supported by global research collaboration
and networking program of Korea Research Council of Fundamental Science \& 
Technology (KRCF) and also partially supported by KASI-Yonsei Joint Research 
Program (2010--2011) for the Frontiers of Astronomy and Space Science funded 
by the Korea Astronomy and Space Science Institute.

  This study critically depends on 43~GHz VLBA data from the Boston University 
gamma-ray blazar monitoring program\footnote{See \web{http://www.bu.edu/blazars/VLBAproject.html}}
\citep{r:bu2} funded by NASA through the Fermi Guest Investigator Program. 
We thank Alan Marscher and Svetlana Jorstad for making results 
of their analysis in the form of brightness distributions publicly available 
on-line prior publication. The Very Long Baseline Array is an instrument of the 
National Radio Astronomy Observatory (NRAO). NRAO is a facility of the 
National Science Foundation, operated by Associated Universities Inc.

{\it Facilities:} \facility{Korean VLBI Network}

\end{document}